\documentclass[twocolumn,twoside,slac_two]{revtex4}
\usepackage{graphicx}
\usepackage{epsfig}
\usepackage{fancyhdr}
\begin{document}

\title{Theoretical Status of Higgs Production at Hadron Colliders in 
the Standard Model}

\author{Radja Boughezal}
\affiliation{Institut f\"ur Theoretische Physik,
        Universit\"at Z\"urich,\\
        Winterthurerstr. 190,
        8057 Z\"urich, Switzerland}

\begin{abstract}
We briefly review the current status of theoretical calculations for Higgs 
production at hadron colliders within the Standard Model. We focus
on the main production mechanisms and decay modes at the Tevatron and the LHC.
\end{abstract}

\maketitle

\thispagestyle{fancy}


\section{Introduction}
%
The search for the Higgs boson, the last missing particle in 
the Standard Model (SM) responsible for the electroweak symmetry breaking, 
is a primary goal of the CERN Large Hadron Collider 
(LHC), and is a central part of Fermilab's Tevatron program. 
In the Standard Model, mass generation is triggered by the Higgs mechanism,
which predicts the existence of one scalar particle, 
the Higgs boson~\cite{{Higgs:1964pj}}.
The coupling of the Higgs to fermions and gauge bosons is predicted by the
model. The only unknown parameter is the Higgs boson mass. 
Direct searches at LEP restrict the Higgs boson mass to be greater 
than 114.4 GeV (at 95$\%$ CL)~\cite{Barate:2003sz}, while precision 
measurements point to a rather light Higgs, 
$M_H \leq 157$~GeV ($95\%$ CL) which increases to
$186$~GeV when including the LEPII direct search limit of $114$ GeV 
(see~\cite{LEPweb} for regular updates). Recently, the Tevatron collaborations,
CDF and D0, reported a $95\%$ CL exclusion of a Standard Model Higgs boson
mass in the range $160 < M_H < 170$ GeV~\cite{:2009pt}. \\
The Standard Model Higgs coupling is strongest to the heaviest
particles. Therefore, we distinguish three types of decays: into fermions,
into massive gauge bosons and loop-induced decays through a massive loop 
of quarks or gauge bosons. 
Since the LHC will be able to find the Higgs, if it exists,
and can provide a measurement of its couplings at the $10-30\%$ 
level~\cite{Duhrssen:2004cv,Lafaye:2009vr}, precise theoretical 
predictions of these decays are needed. \\
Understanding the theoretical prediction is crucial to both the search for and
exclusion of the Standard Model Higgs boson.  Backgrounds to the Higgs signal
are severe in many channels, particularly when a mass peak cannot be
reconstructed such as in~$H\to WW \to l\nu l\nu$, 
and knowledge of the signal shape and normalization
is needed to optimize experimental searches. Signal and background cross 
sections must be, therefore, predicted as accurately as can be achieved.\\
Higgs production at both hadron colliders, Tevatron and LHC, is dominated by
gluon fusion, where two incoming gluons produce a Higgs boson via a virtual
top quark loop. This is followed by vector boson fusion (VBF), where 
the incoming protons radiate a W or a Z boson, which subsequently interact
weakly and fuse into a Higgs boson. The Higgs can also be produced in
association with a pair of top quarks or through Higgs strahlung (associated
WH or ZH production).\\

In this short review, we summarize the status of theoretical predictions
for signal and background processes at hadron colliders. Readers who are
interested in more details should refer to several reviews in the
literature, for example~\cite{Djouadi:2005gi,Harlander:2007zz}.      
%
%
\section{Higgs Decay Modes}
\begin{figure}[t]
 \begin{center}
   \hspace*{-.4cm}\includegraphics[width=0.38\textwidth, angle=-90]{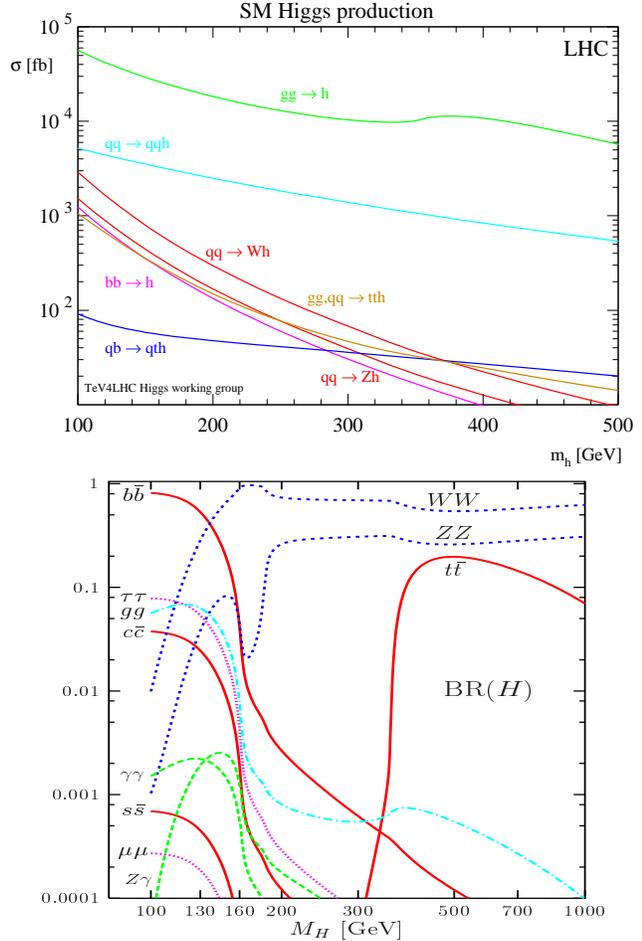}
     \hspace*{-.3cm}\includegraphics[width=.92\linewidth,height=0.75\linewidth]{bra.epsi}
 \caption{Total cross sections for Higgs production 
     at the LHC (upper picture) and branching ratios for a Standard Model 
     Higgs boson~\cite{Duperrin:2008in}(lower picture).
\vspace*{-1.4cm}
}
\label{fig:SMHPRODDEC}
 \end{center}
\end{figure}

%
Depending on the Higgs boson mass, different decay channels open up as
shown in Fig.~\ref{fig:SMHPRODDEC}. The main decay modes are summarized 
below.\\\\
$\bf{H\to b\bar{b}}$:
at low Higgs mass ($M_H \leq 130$ GeV), the decay to $b\bar{b}$ is dominant
with a branching ratio (BR) of roughly $90\%$ for Higgs masses lower 
than $100$ GeV.
Electroweak corrections to this decay were calculated 
at the one- and two-loop level 
in~\cite{Fleischer:1980ub,Kniehl:1991ze,Dabelstein:1991ky,Djouadi:1997rj}.  
The one-loop correction grows like $G_F m_t^2$ and has an impact of $0.3\%$ 
with respect to the LO. 
Mixed QCD-electroweak corrections of order 
$\mathcal O (\alpha_s G_F m_t^2)$ and $\mathcal O (\alpha_s^2 G_F m_t^2)$
were provided
in~\cite{Kwiatkowski:1994cu,Kniehl:1994ju,Kniehl:1995at,Kniehl:1995br,Chetyrkin:1996ke,Chetyrkin:1996wr}. They
were found to be very small, roughly $-0.24\%$ at $O(\alpha_s G_F m_t^2)$. 
Pure QCD corrections to decays into quarks were considered up to $\mathcal O
(\alpha_s^4)$ and are presented in~\cite{Baikov:2005rw}. 
They increase the leading order by $25\%$.\\\\
\noindent
  $ \bf{H \to \tau \tau}$: this is the second important decay at low Higgs mass
after the H$\to b\bar{b}$ with a branching ratio of roughly $10\%$.
The CDF collaboration conducted recently a search for the Higgs using
this decay mode~\cite{CDF:08,Duperrin:2008in}. 
Several processes have been considered: Higgs production
in association with a vector boson (W/Z) with the vector boson decaying into two
jets, vector boson fusion production in which the two jets coming from
the proton and antiproton tend to have a large rapidity value, and gluon fusion
production. This decay is, for the first time, included in the Tevatron 
combined results shown in~\cite{Group:2008ds}.
\\\\
\noindent
 $ \bf{H \to W^+W^-/ZZ}$: with increasing mass, 
the Higgs boson decays preferably
to the heaviest particles, mainly to $W^+W^-$ and ZZ pairs around 
their thresholds (BR $\approx 98\%$ around $M_H=160$ GeV for the WW). 
In the range $140 < M_H < 180$ GeV, 
the $W^+W^- \to l^+l^-\nu\bar{\nu}$ decay is the most important.
Because of the missing energy, the mass of the Higgs can not be directly 
reconstructed and a mass peak is absent. The charged leptons, however, have 
a strong angular correlation.\\ \\ 
Radiative corrections of the strong and electroweak interactions 
at NLO were calculated for the Higgs boson decay $H \to W^+W^- /ZZ \to 4 f$
with semi-leptonic or hadronic four-fermion final states 
in ~\cite{Bredenstein:2006ha}, whereas the pure leptonic final state was
considered in~\cite{Bredenstein:2006rh}. 
The electroweak corrections
are similar for all four-fermion final states and reach $7-8\%$ 
at $M_H \sim 500$GeV. The QCD corrections to the partial decay widths 
are $3.8\%$ for semi-leptonic and $7.6\%$ for hadronic final states.
\\\\
The impact of higher-order corrections on the Higgs signal is strongly reduced 
by the selection cuts that are imposed to suppress the $t\bar{t}$
background~\cite{Anastasiou:2007mz,Grazzini:2008tf,Anastasiou:2008ik}. We
will comment on this in more detail in section~(\ref{SEC:ggh}). \\\\
\noindent
The $W^+W^- \to l^+l^-\nu\bar{\nu}$ is considered to be one of the most 
promising
channels for an early discovery~\cite{Davatz:2004zg}, but at the same time 
a very challenging one due to the background. Therefore, a precise knowledge
of the background distributions is crucial. The irreducible $W^+W^-$ and $ZZ$ 
backgrounds are known at 
$\mathcal O (\alpha_s)$ including spin 
correlations~\cite{Dixon:1999di,Campbell:1999ah}. In the $WW$ case, 
NLO predictions were consistently combined 
with soft-gluon resummation that is valid at small transverse momenta 
of the WW pair~\cite{Grazzini:2005vw}, whereas for ZZ, soft-gluon effects
on signal and background were studied recently in~\cite{Frederix:2008vb}.
The $t\bar{t}$ background, including
spin correlations~\cite{Bernreuther:2001rq}, is known up to NLO in the QCD
coupling. The background from $gg \to W^+W^- \to 4$ leptons is known
at $\mathcal O (\alpha_s^2)$~\cite{Binoth:2005ua} (note that this is NLO
precision as the leading order process is already one loop) and was found 
to increase the theoretical background estimate by almost $30\%$. 
The $gg \to Z (\gamma)Z(\gamma) \to l\bar{l}\,{l'}\bar{l'}$ was considered 
in~\cite{Binoth:2008pr}. In~\cite{Chachamis:2008yb}, the two-loop 
and the one-loop squared virtual QCD corrections to the W boson pair 
production in the $q\bar{q}$ channel in the limit where all kinematical 
invariants are large compared to the mass of the W boson are presented.  
\\
The $H \to ZZ^{(*)} \to 4$ leptons decay is the `golden channel' for observing 
a Higgs boson as a clear peak on top of a smooth background~\cite{nisati:09}.
Its branching ratio is roughly half the one for the W-pair as can be seen 
from Fig.~\ref{fig:SMHPRODDEC}.
The advantage of this channel over the WW one, is that the invariant mass 
of the leptons can be reconstructed, allowing a measurement of the background 
from the data.\\\\
$\bf{H \to \gamma\gamma}$: for low Higgs masses, the dominant decay mode
$H \to b\bar{b}$ is swamped by a large QCD background and the Higgs boson can
be searched for through loop induced decays. The $H \to \gamma\gamma$ is
the most important one and is mediated by loops of massive quarks as well
as massive vector bosons. The $\mathcal O (\alpha_s)$ QCD corrections to this
decay are known for arbitrary quark
masses~\cite{Djouadi:1993ji,Fleischer:2004vb,Harlander:2005rq,Passarino:2007fp},and
do not exceed $5\%$. The $\mathcal O (\alpha_s^2)$ term is known 
as an expansion in $M_H^2/M_t^2$ from the work 
of~\cite{Steinhauser:1996wy}. Finally, electroweak corrections to this decay
were evaluated in~\cite{Fugel:2004ug,Degrassi:2005mc,Passarino:2007fp}. 
They were found to be below $4\%$.

\section{Higgs Production Modes}

\subsection{Gluon Fusion}
\label{SEC:ggh}
\begin{figure}[h]
\vspace{0.3cm}
\begin{center}
\includegraphics[width=0.35\textwidth]{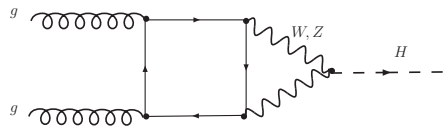}
\caption{{Example two-loop light-quark diagram contributing to the Higgs
boson production cross section via gluon fusion.}}
\label{fig:2loop}
\end{center}
\end{figure}

The dominant production mode of the Standard Model Higgs boson at the Tevatron
and LHC is gluon fusion, mediated by a heavy-quark loop,
with a cross section that is a factor of $10$ larger than 
all other production modes cross sections (see Fig.~(\ref{fig:SMHPRODDEC})). 
Radiative QCD corrections to this process turned out to be very important. 
At the NLO level, they were found to increase the LO cross
section by about $80-100\%$~\cite{Dawson:1990zj,Djouadi:1991tka,Spira:1995rr}.
The gluon-Higgs interaction seems to be very well 
approximated by an effective Lagrangian obtained
by decoupling the top quark~\cite{Kramer:1996iq}

\begin{equation}
{\cal L}_{eff} = -\alpha_s\frac{C_1}{4v} H G_{\mu\nu}^a G^{a\mu\nu},
\label{lag}
\end{equation}
if the exact Born cross section with the full dependence on the top and 
bottom quark masses is used to normalize the result. The difference between
the exact and the approximate NLO cross sections is less than $1\%$
for Higgs masses up to $200$ GeV, and does not exceed $10\%$ 
even for Higgs masses up to $1$~TeV, well far away from its formal range 
of validity $ M_H< 2 M_{top} $. 
In equation~(\ref{lag}), $v$ is the vacuum expectation value of 
the Higgs field, $v=246$~GeV, and $C_1$ is the Wilson coefficient, 
currently known through $\alpha_s^5$~\cite{Schroder:2005hy,Chetyrkin:2005ia}.   
In this large $M_{top}$ limit, the NNLO QCD corrections were 
computed in~\cite{Harlander:2002wh,Anastasiou:2002yz,Ravindran:2003um}, leading
to an additional increase of the cross section of roughly $10-15\%$, and
showing a good convergence of the perturbative series. Very recently, 
the three-loop virtual corrections to this process, where finite top
quark mass effects are taken into account, were presented 
in~\cite{Harlander:2009bw,Pak:2009bx}.\\
A further increase in the cross section of about $6\%$ was obtained by 
doing soft-gluon resummation~\cite{Catani:2003zt}. This result was nicely
confirmed through the leading soft contributions at 
$N^3LO$~\cite{Moch:2005ky,Laenen:2005uz,Idilbi:2005ni}.
Taking all the perturbative effects into account, the inclusive result 
of the cross section increases by a factor of $2$ at LHC and $3.5$ at Tevatron. 
The theoretical uncertainty from effects beyond NNLO is estimated to be about 
$\pm 10 \%$ by varying the renormalization and factorization scales.\\\\
The largeness of the K-factors at both colliders is an open question. 
In~\cite{Ahrens:2008nc}, the authors argue that this is due to enhanced
contributions of the form $(C_A \pi\alpha_s)^L$,  coming from 
the analytic continuation of the gluon form factor to time-like 
momentum transfer, with $L$ being the number of loops. 
A resummation of these terms using soft collinear effective theory (SCET)
leads to smaller values of the K-factors; in fact at the LHC, the K-factor 
for small values of $M_H$ is close to $1$.  \\\\
The importance and success in taming the QCD corrections to Higgs production have shifted attention to electroweak corrections to the Higgs signal.  The authors of Refs.~\cite{Aglietti:2004nj,Aglietti:2006yd} pointed out important 2-loop light-quark effects; these
are pictured in Fig.~(\ref{fig:2loop}) and involve the Higgs coupling to $W$- or $Z$-bosons which then couple to gluons through a light-quark loop. These terms are not suppressed by light-quark Yukawa couplings, and receive a multiplicity enhancement from summing over the quarks.  A careful study of the full 2-loop electroweak effects was performed in Ref.~\cite{Actis:2008ug}. They increase the leading-order cross section by
up to $5-6\%$ for relevant Higgs masses.

\begin{figure}
 \begin{center}
   \includegraphics[width=0.23\textwidth]{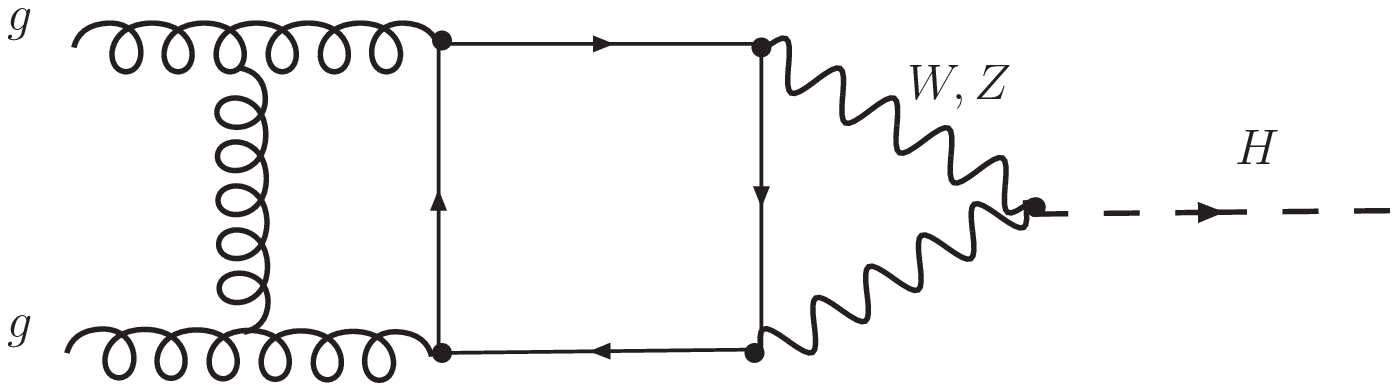}
  \hspace*{-.0cm} \includegraphics[width=0.23\textwidth]{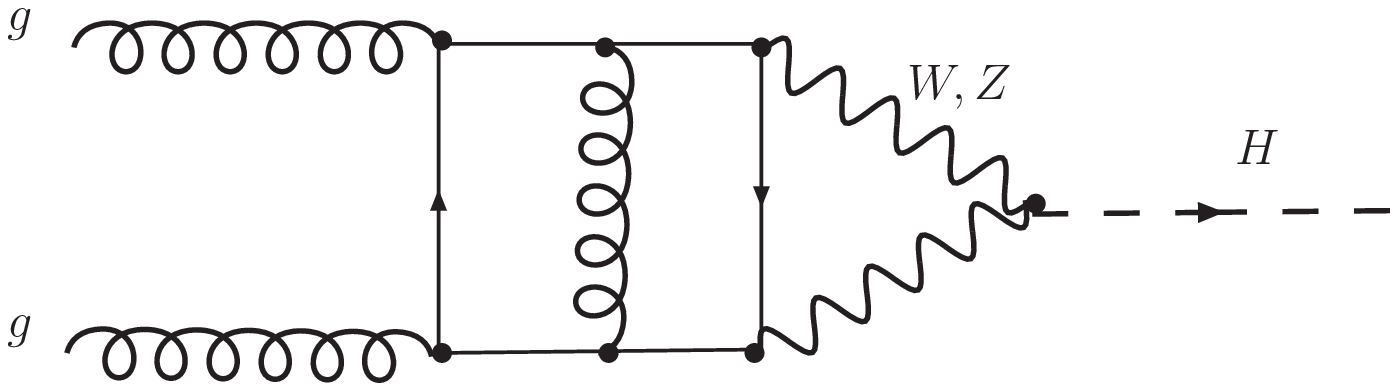}
   \caption{{Example three-loop light-quark diagrams contributing
            to the $\mathcal{O}(\alpha\alpha_s)$ term.}}
   \label{fig:3loop}
 \end{center}
\end{figure}

\begin{figure}
 \begin{center}
   \includegraphics[width=0.5\textwidth]{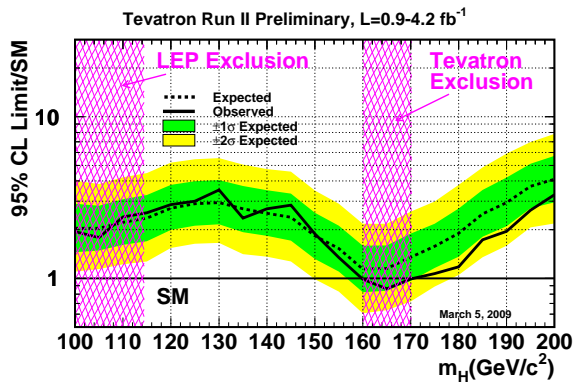}
   \caption{{The Tevatron exclusion limits of a SM Higgs boson mass
    in the range $160-170$ GeV at $95\%$ CL~\cite{:2009pt}.}}
   \label{fig:TevExc09}
 \end{center}
\end{figure}

The leading order of the mixed QCD-electroweak corrections, due to diagrams
containing light quarks, was calculated in~\cite{Anastasiou:2008tj}, 
using an effective field theory approach where the W boson is integrated out.
Sample diagrams involved in this calculation are shown in 
Fig.~(\ref{fig:3loop}).
This work allowed to check the complete factorization hypothesis of QCD and
electroweak corrections proposed in Ref.~\cite{Aglietti:2006yd,Actis:2008ug}.
The result shows that, despite the large violation of the factorization 
assumption,
a significant numerical difference from the prediction of this hypothesis
is not observed in the cross section, due to the dominant QCD corrections.
The end effect on the cross section is an additional enhancement 
of up to $6\%$ from the $\mathcal{O}(\alpha) + \mathcal{O}(\alpha\alpha_s)$ 
terms. \\\\
In addition to the previous results, the authors of~\cite{Anastasiou:2008tj}
provided a new prediction for the inclusive cross section of the gluon
fusion process. This updated result takes into account all the new theoretical
calculations: the 2-loop light-quark diagrams based on the complex-mass scheme 
for the $W$- and $Z$-bosons~\cite{Actis:2008ug}, the new 3-loop 
$\mathcal{O}(\alpha\alpha_s)$ correction,
the contributions from top and bottom quarks with the exact NLO K-factors
and the newest parton distribution functions (PDF) by the MSTW group~\cite{Martin:2009iq,Martin:2009bu,Thorne:2009ky}.
The updated numerical values of the cross section are $4-6\%$ lower than 
the old prediction~\cite{Catani:2003zt} that was used in an earlier 
exclusion of a SM Higgs boson mass of $170$ GeV by the Tevatron. 
Numerical values for the new prediction are shown in 
Table~(\ref{tab:TEVxsec08}). 
We note that similar results were obtained in~\cite{deFlorian:2009hc}.
The new prediction, together with new data collected by the Tevatron 
collaborations, were used to provide a new excluded range
of Higgs masses, namely $160-170$~GeV mass range with $95\%$ CL~\cite{:2009pt}. 
See Fig.~(\ref{fig:TevExc09}).
\begin{table}[t]
  \begin{center}
  \resizebox{0.45\textwidth}{!}{
    \begin{tabular}{|c|c||c|c|}
      \hline
      $m_{H}$[GeV] &$\sigma^{best}$[pb]&$m_{H}$[GeV] &$\sigma^{best}$[pb]\\
        \hline \hline
     110&    1.417 ($\pm 7\%$ pdf) &160&0.4344 ($\pm 9\%$ pdf)
 \\[1mm]
      \hline
  115&    1.243 ($\pm 7\%$ pdf) &165&0.3854 ($\pm 9\%$ pdf)
\\[1mm]
      \hline
  120&    1.094  ($\pm 7\%$ pdf)  &170&0.3444 ($\pm 10\%$ pdf)
\\[1mm]
      \hline
  125 &   0.9669 ($\pm 7\%$ pdf) &175&0.3097 ($\pm 10\%$ pdf)
\\[1mm]
      \hline
  130 &  0.8570 ($\pm 8\%$ pdf) &180&0.2788 ($\pm 10\%$ pdf)
\\[1mm]
      \hline
  135 &   0.7620 ($\pm 8\%$ pdf) &185&0.2510 ($\pm 10\%$ pdf)
\\[1mm]
      \hline
  140 &   0.6794 ($\pm 8\%$ pdf) &190&0.2266 ($\pm 11\%$ pdf)
\\[1mm]
      \hline
  145 &   0.6073 ($\pm 8\%$ pdf) &195&0.2057 ($\pm 11\%$ pdf)
\\[1mm]
      \hline
  150 &   0.5439 ($\pm 9\%$ pdf) &200&0.1874 ($\pm 11\%$ pdf)
  \\[1mm]
      \hline
  155 & 0.4876  ($\pm 9\%$ pdf) & $-$& $-$
      \\\hline
    \end{tabular}
    }
  \end{center}
  \caption{{Higgs production cross
   section (MSTW08) for Higgs mass values relevant for Tevatron, with
   $\mu= \mu_R =\mu_F=M_H/2 $. The total cross section
   $\sigma^{best} =
   \sigma_{QCD}^{NNLO}+\sigma_{EW}^{NNLO}$~\cite{Anastasiou:2008tj}.  The
   theoretical errors PDFs are shown in the Table; the scale variation is
   $^{+7\%}_{-11\%}$, roughly constant as a function of Higgs boson mass.}
    \label{tab:TEVxsec08} }
\end{table}
\\\\
The calculations mentioned above refer to the inclusive cross section,
which means no experimental cuts were imposed.
It was shown in a previous
work~\cite{Anastasiou:2007mz} that 
the impact of higher order corrections on the rate and shape of 
the corresponding distributions may be strongly dependent on the chosen cuts.
The most general higher order prediction for gluon fusion process is available
on the form of partonic NNLO Monte Carlo
programs~\cite{Anastasiou:2005qj,Catani:2007vq} which use the large-$M_{top}$
limit and vanishing b-quark mass. In~\cite{Anastasiou:2007mz}, 
the authors have shown that, while for the process 
$g g \to H \to \gamma \gamma$, radiative corrections are only slightly
affected by the signal cuts, the process $g g \to H \to WW \to l \nu
\bar{l}\nu$ is strongly affected by these cuts which take away most of the
increase observed in the inclusive cross section. The numbers in tables
Table~(\ref{tab:kfactors}) and Table~(\ref{tab:davatzcuts}) reflect that.    

\begin{table}[htb]
\begin{center}
$$
\begin{array}{||c|c|c||}
\hline m_h,~{\rm GeV} & \sigma^{\rm cut}_{\rm NNLO}/\sigma^{\rm inc}_{\rm NNLO}
 & K^{(2)}_{\rm cut}/K^{(2)}_{\rm inc} \\ \cline{1-3}
110&  0.590 & 0.981 \\ \cline{2-3}
115&  0.597 & 0.968 \\ \cline{2-3}
120&  0.603 & 0.953 \\ \cline{2-3}
125&  0.627 & 0.970 \\ \cline{2-3}
130&  0.656 & 1.00  \\ \cline{2-3}
135&  0.652 & 0.98  \\ \cline{2-3}
\hline
\end{array}
$$
\vspace*{0.5cm}
\caption{\label{tab:kfactors}
Comparisons between the cut and inclusive cross sections for 
$gg \to H \to\gamma\gamma + X$ for different Higgs masses.  
The second column contains the ratio of the NNLO cross section
with the standard cuts over the inclusive cross section, while the third
column contains the ratio of cut and inclusive results for the $K$-factor 
$K^{(2)} = \sigma_{\rm NNLO} /  \sigma_{\rm NLO}$. $\mu_R = \mu_F = M_H/2$.
 See~\cite{Anastasiou:2005qj} for more details.
}
\end{center}
\end{table}

\begin{table}[h]
\begin{center}
\begin{tabular}{||c|c|c|c||}
\hline
$\sigma(\mathrm{fb})$         & LO     & NLO    &   NNLO  \\
\hline
$ \mu=\frac{M_H}{2}$ & $21.002 \pm 0.021$
                   & $22.47 \pm 0.11$
                   & $18.45 \pm 0.54$
                   \\
\hline
$\mu= M_H$         & $17.413 \pm 0.017$
                   & $21.07 \pm 0.11$
                   & $18.75 \pm 0.37$
                   \\
\hline
$\mu= 2 M_H$       & $14.529 \pm 0.014$
                   & $19.50 \pm 0.10$
                   & $19.01 \pm 0.27$
                   \\
\hline
\end{tabular}
\end{center}
\caption{
\label{tab:davatzcuts}
Cross-section through NNLO for the
$gg \to H \to WW \to l\bar{\nu}\bar{l}\nu$ after applying signal cuts.
 See~\cite{Anastasiou:2007mz} for details.}
\end{table}
Very recently, finite top- and bottom-quark mass effects 
and electroweak contributions to the Higgs boson transverse
momentum spectrum ($P_T$-spectrum) at the NLO level
were presented in~\cite{Keung:2009bs,Anastasiou:2009kn}.
\subsection{Vector Boson Fusion}
\begin{figure}
 \begin{center}
   \includegraphics[width=0.35\textwidth]{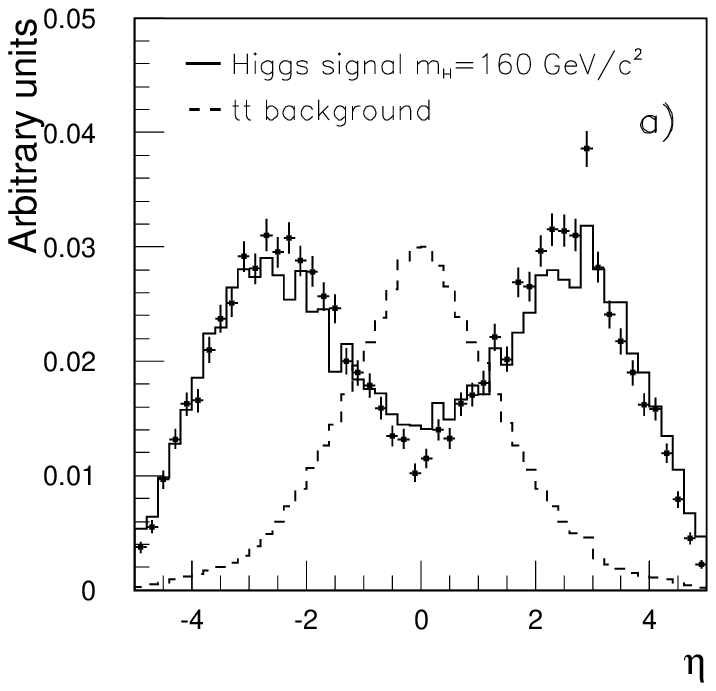}
   \includegraphics[width=0.35\textwidth]{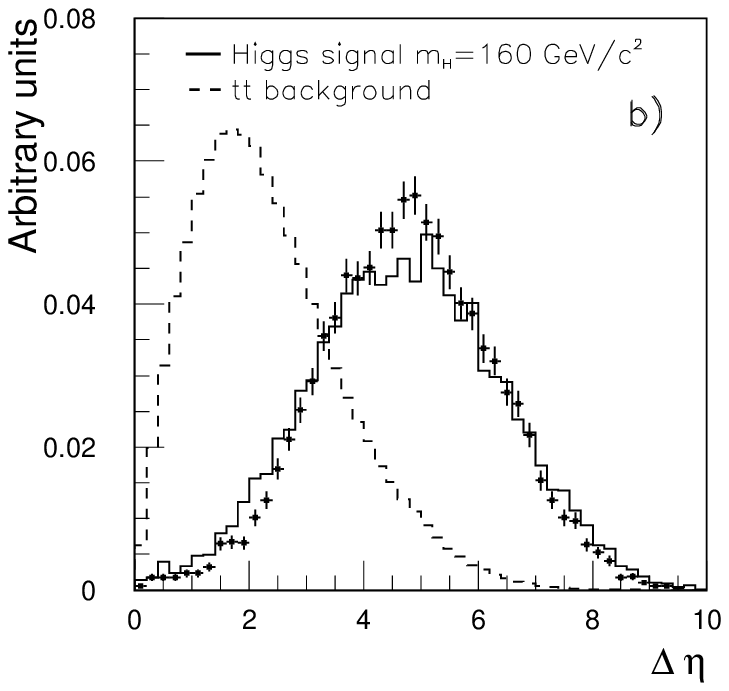}
   \caption{ a) Pseudorapidity distribution of the two tagging jets
in signal events with $m_H = 160$~GeV and for $t\bar{t}$ background events.
b) Rapidity separation between the tagging jets~\cite{Asai:2004ws}.}
   \label{f:tagjets}
 \end{center}
\end{figure}

This process is important for the discovery of the Higgs boson 
at the LHC for a wide range of Higgs masses~\cite{Rainwater:1997dg,Rainwater:1998kj,Plehn:1999xi,Eboli:2000ze}. The vector boson fusion (VBF) 
cross section is one order of magnitude lower than the one for gluon fusion,
but it is an attractive channel for measurements of the Higgs 
couplings and CP
properties~\cite{Zeppenfeld:2000td,Belyaev:2002ua,Duhrssen:2004cv,Klamke:2007cu}. 
In VBF, the Higgs is produced in association with two jets which are scattered
into the forward direction. These two jets are not color connected at LO,
which means that the hadronic activity in the rapidity region between these
two jets is very small. On the other hand, the Higgs decay products are found 
at central rapidities, which allows to efficiently reduce the background 
if suitable cuts are chosen (see Fig.~\ref{f:tagjets}). \\
The NLO QCD corrections to the total cross section were computed a long time ago
and found to be of the order $5-10\%$~\cite{Han:1992hr}. 
These corrections have been implemented in fully differential 
partonic NLO Monte Carlo programs in~\cite{Figy:2003nv,Figy:2004pt,Berger:2004pca}.
The full electroweak and QCD corrections to this process
have also been computed~\cite{Ciccolini:2007jr,Ciccolini:2007ec}.
Like other production modes, the VBF process suffers from a large background.
The dominant one when trying to isolate the $HWW$ and $HZZ$ couplings
through VBF is the Higgs production plus two jets from gluon-gluon
fusion. The LO contribution to this background is known keeping the full
top-mass dependence~\cite{DelDuca:2001fn}. The authors have shown that 
the VBF cuts on 
the rapidity and the transverse momentum of the tagging jets work efficiently 
in this case. The NLO QCD corrections to $Hjj$ in the large top quark mass
limit are also known~\cite{Campbell:2006xx}, as well as parton shower effects
on the azimuthal angle correlation of the two jets and the rapidity 
distribution of extra jets~\cite{DelDuca:2006hk}. 

\subsection{Higgs Strahlung}
\begin{figure}
 \begin{center}
     \hspace{-.7cm}\includegraphics[width=.83\linewidth,height=0.75\linewidth]{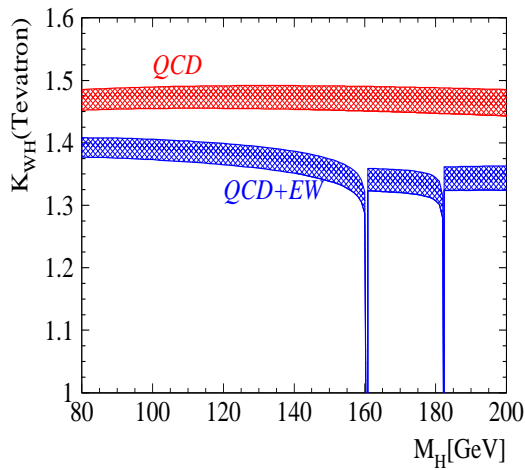}
 \caption{K-factor for WH production at the Tevatron after inclusion of the
NNLO QCD and electroweak $\mathcal{O}(\alpha)$ corrections. 
Theoretical errors are estimated by varying the renormalization 
and factorization scales in the range $1/3 M_{VH} < \mu_R (\mu_F) < 3M_{VH}$, 
the other scale being fixed at $\mu_F (\mu_R) = M_{VH}$. $M_{VH}$ is 
the invariant mass of the VH system~\cite{Brein:2004ue}.}
    \label{fig:WH}
 \end{center}
\end{figure}

The associated production of a Higgs with a W or a Z boson is an 
important discovery channel at Tevatron for a low Higgs mass. It utilizes
the $H \to b\bar{b}$ decay mode and the leptonic decay of the vector
boson to reject the background. It has a small cross section that  
ranges between $0.3$ pb and $3$ pb depending on the Higgs mass.
A recent analysis has shown that a  
signal in this channel might be observable at the LHC despite the  
large backgrounds~\cite{Butterworth:2008iy}.\\
The NLO QCD corrections to this production mode are known~\cite{Han:1991ia}. 
They increase the cross section by
about $30\%$. The NNLO QCD results are also available and give a further
enhancement of the cross section by about 
$5-10\%$~\cite{Brein:2003wg}. These corrections lead to a reduction
of the scale dependence of the cross section from 
$10\%$ at LO to $5\%$ at NLO,  to $2\%$
when the NNLO result is included. At this level of precision, electroweak 
corrections become important to further improve the precision of 
the prediction. They were     
calculated at order $\mathcal{O}(\alpha)$ in~\cite{Ciccolini:2003jy} and were
found to decrease the cross section by $5\%$ to $10\%$ depending on the 
Higgs boson mass and the input parameters scheme. These two types of
corrections were combined to produce an up-to-date cross section. 
The WH K-factor
for the Tevatron is shown as an example 
in Fig.~(\ref{fig:WH})~\cite{Brein:2004ue}. 
\subsection{Associated Production With a $\bf{t\bar{t}}$ Pair}
This channel offers the possibility of measuring the top Yukawa
coupling~\cite{Belyaev:2002ua,Duhrssen:2004cv}. 
It was initially thought to be an important discovery channel in the
low Higgs mass region, by looking at the $H \to b\bar{b}$ decay mode 
and triggering on the leptonic decay of one of the top quarks. The signature
is four b-quarks in association with two W bosons. However, any Higgs decay
product will essentially be present in the top decays, therefore, there
are large backgrounds, particularly from $t\bar{t}b\bar{b}$ and $t\bar{t}jj$.
A detailed analysis of the backgrounds together with a full detector
simulation showed that it is very difficult to observe the Higgs in this 
channel~\cite{Ball:2007zza}.\\
The NLO QCD corrections to this decay are available from the work of two
independent 
groups~\cite{Beenakker:2001rj,Beenakker:2002nc,Dawson:2002tg,Dawson:2003zu},
and turned out to increase the signal cross section by almost $20\%$
at the central scale of $\mu_F=\mu_R=M_t+M_H/2$.
A significant reduction of the renormalization and factorization scales
in the cross section was observed. The NLO QCD background 
$pp \to t\bar{t}H \to t \bar{t}b\bar{b}$ is known from 
the work of~\cite{Bredenstein:2009aj,Bevilacqua:2009zn}.
These corrections significantly reduce the unphysical scale dependence 
of the leading-order cross section, and predict an enhancement 
of the $t \bar{t}b\bar{b}$ production cross section by a K-factor 
of about $1.8$. 
\subsection{Conclusions}

The precision of theoretical calculations has reached an unprecedented 
accuracy. Together with the discovery potential of LHC for 
a SM Higgs boson, we are ready to observe a first evidence of a signal
of this particle, should it exist at all.
We have reviewed the current theoretical status of the most important 
production and decay modes of a Standard Model Higgs boson 
at hadron colliders. New theoretical calculations and predictions for 
production modes, in particular the gluon fusion process, were briefly 
discussed.
Further more, new results for background processes to Higgs boson 
production, like the $t\bar{t}H$, were also sketched.    


\begin{acknowledgments}
This work is supported by the Swiss National Science Foundation
under contract 200020-116756/2.
\end{acknowledgments}


\bigskip 

%


\begin{thebibliography}{100}
%
%
\bibitem{Higgs:1964pj}
  P.~W.~Higgs,
  Phys.\ Rev.\ Lett.\  {\bf 13} (1964) 508.
%
\bibitem{Barate:2003sz}
  R.~Barate {\it et al.}  [LEP Working Group for Higgs boson searches and
                  ALEPH Collaboration and  and],
  Phys.\ Lett.\  B {\bf 565} (2003) 61
  [arXiv:hep-ex/0306033].
%
\bibitem{LEPweb}
Home page of LEP Electroweak Working Group: 
{\it http://lepewwg.web.cern.ch/LEPEWWG/}
%
\bibitem{:2009pt}
    [CDF Collaboration and D0 Collaboration],
  arXiv:0903.4001 [hep-ex].
%
\bibitem{Duhrssen:2004cv}
  M.~Duhrssen, S.~Heinemeyer, H.~Logan, D.~Rainwater, G.~Weiglein and D.~Zeppenfeld,
  Phys.\ Rev.\  D {\bf 70} (2004) 113009
  [arXiv:hep-ph/0406323].
\bibitem{Lafaye:2009vr}
  R.~Lafaye, T.~Plehn, M.~Rauch, D.~Zerwas and M.~Duhrssen,
  arXiv:0904.3866 [hep-ph].
%
%
\bibitem{Djouadi:2005gi}
  A.~Djouadi,
  Phys.\ Rept.\  {\bf 457} (2008) 1
  [arXiv:hep-ph/0503172].
%
\bibitem{Harlander:2007zz}
  R.~Harlander,
  J.\ Phys.\ G {\bf 35} (2008) 033001.
%
\bibitem{Fleischer:1980ub}
  J.~Fleischer and F.~Jegerlehner,
  Phys.\ Rev.\  D {\bf 23} (1981) 2001.
%
\bibitem{Kniehl:1991ze}
  B.~A.~Kniehl,
  Nucl.\ Phys.\  B {\bf 376} (1992) 3.
%
\bibitem{Dabelstein:1991ky}
  A.~Dabelstein and W.~Hollik,
  Z.\ Phys.\  C {\bf 53} (1992) 507.
%
\bibitem{Djouadi:1997rj}
  A.~Djouadi, P.~Gambino and B.~A.~Kniehl,
  Nucl.\ Phys.\  B {\bf 523} (1998) 17
  [arXiv:hep-ph/9712330].
%
\bibitem{Kwiatkowski:1994cu}
  A.~Kwiatkowski and M.~Steinhauser,
  Phys.\ Lett.\  B {\bf 338} (1994) 66
  [Erratum-ibid.\  B {\bf 342} (1995) 455]
  [arXiv:hep-ph/9405308].
%
\bibitem{Kniehl:1994ju}
  B.~A.~Kniehl and M.~Spira,
  Nucl.\ Phys.\  B {\bf 432} (1994) 39
  [arXiv:hep-ph/9410319].
%
\bibitem{Kniehl:1995at}
  B.~A.~Kniehl and M.~Steinhauser,
  Nucl.\ Phys.\  B {\bf 454} (1995) 485
  [arXiv:hep-ph/9508241].
%
\bibitem{Kniehl:1995br}
  B.~A.~Kniehl and M.~Steinhauser,
  Phys.\ Lett.\  B {\bf 365} (1996) 297
  [arXiv:hep-ph/9507382].
%
\bibitem{Chetyrkin:1996ke}
  K.~G.~Chetyrkin, B.~A.~Kniehl and M.~Steinhauser,
  Nucl.\ Phys.\  B {\bf 490} (1997) 19
  [arXiv:hep-ph/9701277].
%
\bibitem{Chetyrkin:1996wr}
  K.~G.~Chetyrkin, B.~A.~Kniehl and M.~Steinhauser,
  Phys.\ Rev.\ Lett.\  {\bf 78} (1997) 594
  [arXiv:hep-ph/9610456].
%
\bibitem{Baikov:2005rw}
  P.~A.~Baikov, K.~G.~Chetyrkin and J.~H.~Kuhn,
  Phys.\ Rev.\ Lett.\  {\bf 96} (2006) 012003
  [arXiv:hep-ph/0511063].
%
\bibitem{CDF:08}
CDF Note 9248, \emph{Search for the SM Higgs Boson
using $\tau$ lepton. Simultaneous Search for}
WH/ZH/VBF/ggH \emph{in 2t's+2jets event}, (2008).
%
\bibitem{Duperrin:2008in}
  A.~Duperrin,
  Eur.\ Phys.\ J.\  C {\bf 59} (2009) 297
  [arXiv:0805.3624 [hep-ex]].
%
\bibitem{Group:2008ds}
  T.~T.~W.~Group  [CDF Collaboration and D0 Collaboration],
  arXiv:0804.3423 [hep-ex].
%
\bibitem{Bredenstein:2006ha}
  A.~Bredenstein, A.~Denner, S.~Dittmaier and M.~M.~Weber,
  JHEP {\bf 0702} (2007) 080
  [arXiv:hep-ph/0611234].
%
\bibitem{Bredenstein:2006rh}
  A.~Bredenstein, A.~Denner, S.~Dittmaier and M.~M.~Weber,
  Phys.\ Rev.\  D {\bf 74} (2006) 013004
  [arXiv:hep-ph/0604011].
%
\bibitem{Anastasiou:2007mz}
  C.~Anastasiou, G.~Dissertori and F.~Stockli,
  JHEP {\bf 0709} (2007) 018
  [arXiv:0707.2373 [hep-ph]].
%
\bibitem{Grazzini:2008tf}
  M.~Grazzini,
  JHEP {\bf 0802} (2008) 043
  [arXiv:0801.3232 [hep-ph]].
%
\bibitem{Anastasiou:2008ik}
  C.~Anastasiou, G.~Dissertori, F.~Stockli and B.~R.~Webber,
  JHEP {\bf 0803} (2008) 017
  [arXiv:0801.2682 [hep-ph]].
%
\bibitem{Davatz:2004zg}
  G.~Davatz, G.~Dissertori, M.~Dittmar, M.~Grazzini and F.~Pauss,
  JHEP {\bf 0405} (2004) 009
  [arXiv:hep-ph/0402218].
%
\bibitem{Dixon:1999di}
  L.~J.~Dixon, Z.~Kunszt and A.~Signer,
  Phys.\ Rev.\  D {\bf 60} (1999) 114037
  [arXiv:hep-ph/9907305].
%
\bibitem{Campbell:1999ah}
  J.~M.~Campbell and R.~K.~Ellis,
  Phys.\ Rev.\  D {\bf 60} (1999) 113006
  [arXiv:hep-ph/9905386].
%
\bibitem{Grazzini:2005vw}
  M.~Grazzini,
  JHEP {\bf 0601} (2006) 095
  [arXiv:hep-ph/0510337].
%
\bibitem{Frederix:2008vb}
  R.~Frederix and M.~Grazzini,
  Phys.\ Lett.\  B {\bf 662} (2008) 353
  [arXiv:0801.2229 [hep-ph]].
%
\bibitem{Bernreuther:2001rq}
  W.~Bernreuther, A.~Brandenburg, Z.~G.~Si and P.~Uwer,
  Phys.\ Rev.\ Lett.\  {\bf 87} (2001) 242002
  [arXiv:hep-ph/0107086].
%
\bibitem{Binoth:2005ua}
  T.~Binoth, M.~Ciccolini, N.~Kauer and M.~Kramer,
  JHEP {\bf 0503} (2005) 065
  [arXiv:hep-ph/0503094].
%
\bibitem{Binoth:2008pr}
  T.~Binoth, N.~Kauer and P.~Mertsch,
  arXiv:0807.0024 [hep-ph].
%
\bibitem{Chachamis:2008yb}
  G.~Chachamis, M.~Czakon and D.~Eiras,
  JHEP {\bf 0812} (2008) 003
  [arXiv:0802.4028 [hep-ph]].
%
\bibitem{nisati:09}
See the talk by A. Nisati ``\emph{Anticipating at the LHC}'', 
KITP 2008,\\{\small{{$\tt http://online.itp.ucsb.edu/online/lhc_c08/nisati$}}}
%
\bibitem{Djouadi:1993ji}
  A.~Djouadi, M.~Spira and P.~M.~Zerwas,
  Phys.\ Lett.\  B {\bf 311} (1993) 255
  [arXiv:hep-ph/9305335].
%
\bibitem{Fleischer:2004vb}
  J.~Fleischer, O.~V.~Tarasov and V.~O.~Tarasov,
  Phys.\ Lett.\  B {\bf 584} (2004) 294
  [arXiv:hep-ph/0401090].
%
\bibitem{Harlander:2005rq}
  R.~Harlander and P.~Kant,
  JHEP {\bf 0512} (2005) 015
  [arXiv:hep-ph/0509189].
%
\bibitem{Passarino:2007fp}
  G.~Passarino, C.~Sturm and S.~Uccirati,
  Phys.\ Lett.\  B {\bf 655} (2007) 298
  [arXiv:0707.1401 [hep-ph]].
%
\bibitem{Steinhauser:1996wy}
  M.~Steinhauser,
  arXiv:hep-ph/9612395.
%
\bibitem{Fugel:2004ug}
  F.~Fugel, B.~A.~Kniehl and M.~Steinhauser,
  Nucl.\ Phys.\  B {\bf 702} (2004) 333
  [arXiv:hep-ph/0405232].
%
\bibitem{Degrassi:2005mc}
  G.~Degrassi and F.~Maltoni,
  Nucl.\ Phys.\  B {\bf 724} (2005) 183
  [arXiv:hep-ph/0504137].
%
%
%
\bibitem{Dawson:1990zj}
  S.~Dawson,
  Nucl.\ Phys.\  B {\bf 359}, 283 (1991).
\bibitem{Djouadi:1991tka}
  A.~Djouadi, M.~Spira and P.~M.~Zerwas,
  Phys.\ Lett.\  B {\bf 264}, 440 (1991).
%
\bibitem{Spira:1995rr}
D.~Graudenz, M.~Spira and P.~M.~Zerwas,
  Phys.\ Rev.\ Lett.\  {\bf 70}, 1372 (1993);
  M.~Spira, A.~Djouadi, D.~Graudenz and P.~M.~Zerwas,
  Nucl.\ Phys.\  B {\bf 453}, 17 (1995)
  [arXiv:hep-ph/9504378].
%
\bibitem{Kramer:1996iq}
  M.~1.~Kramer, E.~Laenen and M.~Spira,
  Nucl.\ Phys.\  B {\bf 511} (1998) 523
  [arXiv:hep-ph/9611272].
%
\bibitem{Schroder:2005hy}
  Y.~Schroder and M.~Steinhauser,
  JHEP {\bf 0601} (2006) 051
  [arXiv:hep-ph/0512058].
%
\bibitem{Chetyrkin:2005ia}
  K.~G.~Chetyrkin, J.~H.~Kuhn and C.~Sturm,
  Nucl.\ Phys.\  B {\bf 744} (2006) 121
  [arXiv:hep-ph/0512060].
%
\bibitem{Harlander:2002wh}
  R.~V.~Harlander and W.~B.~Kilgore,
  Phys.\ Rev.\ Lett.\  {\bf 88}, 201801 (2002)
  [arXiv:hep-ph/0201206].

\bibitem{Anastasiou:2002yz}
  C.~Anastasiou and K.~Melnikov,
  Nucl.\ Phys.\  B {\bf 646}, 220 (2002)
  [arXiv:hep-ph/0207004].
\bibitem{Ravindran:2003um}
  V.~Ravindran, J.~Smith and W.~L.~van Neerven,
  Nucl.\ Phys.\  B {\bf 665}, 325 (2003)
  [arXiv:hep-ph/0302135].
%
\bibitem{Harlander:2009bw}
  R.~V.~Harlander and K.~J.~Ozeren,
  arXiv:0907.2997 [hep-ph].
%
\bibitem{Pak:2009bx}
  A.~Pak, M.~Rogal and M.~Steinhauser,
  arXiv:0907.2998 [hep-ph].
%
\bibitem{Catani:2003zt}
  S.~Catani, D.~de Florian, M.~Grazzini and P.~Nason,
  JHEP {\bf 0307} (2003) 028
  [arXiv:hep-ph/0306211].
%
\bibitem{Moch:2005ky}
  S.~Moch and A.~Vogt,
  Phys.\ Lett.\  B {\bf 631} (2005) 48
  [arXiv:hep-ph/0508265].
%
\bibitem{Laenen:2005uz}
  E.~Laenen and L.~Magnea,
  Phys.\ Lett.\  B {\bf 632} (2006) 270
  [arXiv:hep-ph/0508284].
%
\bibitem{Idilbi:2005ni}
  A.~Idilbi, X.~d.~Ji, J.~P.~Ma and F.~Yuan,
  Phys.\ Rev.\  D {\bf 73} (2006) 077501
  [arXiv:hep-ph/0509294].
%
\bibitem{Ahrens:2008nc}
  V.~Ahrens, T.~Becher, M.~Neubert and L.~L.~Yang,
  Eur.\ Phys.\ J.\  C {\bf 62} (2009) 333
  [arXiv:0809.4283 [hep-ph]].
%
\bibitem{Aglietti:2004nj}
  U.~Aglietti, R.~Bonciani, G.~Degrassi and A.~Vicini,
  Phys.\ Lett.\  B {\bf 595}, 432 (2004)
  [arXiv:hep-ph/0404071].
%
\bibitem{Aglietti:2006yd}
  U.~Aglietti, R.~Bonciani, G.~Degrassi and A.~Vicini,
  arXiv:hep-ph/0610033.
%
\bibitem{Actis:2008ug}
  S.~Actis, G.~Passarino, C.~Sturm and S.~Uccirati,
  arXiv:0809.1301 [hep-ph];
S.~Actis, G.~Passarino, C.~Sturm and S.~Uccirati,
  arXiv:0809.3667 [hep-ph].
%
\bibitem{Anastasiou:2008tj}
  C.~Anastasiou, R.~Boughezal and F.~Petriello,
  JHEP {\bf 0904} (2009) 003
  [arXiv:0811.3458 [hep-ph]].
%
\bibitem{Martin:2009iq}
  A.~D.~Martin, W.~J.~Stirling, R.~S.~Thorne and G.~Watt,
  arXiv:0901.0002 [hep-ph]
%
  \bibitem{Martin:2009bu}
  A.~D.~Martin, W.~J.~Stirling, R.~S.~Thorne and G.~Watt,
  arXiv:0905.3531 [hep-ph]
%
  \bibitem{Thorne:2009ky}
  R.~S.~Thorne, A.~D.~Martin, W.~J.~Stirling and G.~Watt,
  arXiv:0907.2387 [hep-ph].
%
\bibitem{deFlorian:2009hc}
  D.~de Florian and M.~Grazzini,
  Phys.\ Lett.\  B {\bf 674} (2009) 291
  [arXiv:0901.2427 [hep-ph]].
%
\bibitem{Anastasiou:2005qj}
  C.~Anastasiou, K.~Melnikov and F.~Petriello,
  Nucl.\ Phys.\  B {\bf 724} (2005) 197
  [arXiv:hep-ph/0501130].
%
\bibitem{Catani:2007vq}
  S.~Catani and M.~Grazzini,
  Phys.\ Rev.\ Lett.\  {\bf 98} (2007) 222002
  [arXiv:hep-ph/0703012].
%
\bibitem{Keung:2009bs}
  W.~Y.~Keung and F.~J.~Petriello,
  Phys.\ Rev.\  D {\bf 80} (2009) 013007
  [arXiv:0905.2775 [hep-ph]].
%
\bibitem{Anastasiou:2009kn}
  C.~Anastasiou, S.~Bucherer and Z.~Kunszt,
  arXiv:0907.2362 [hep-ph].
%
%
%
\bibitem{Zeppenfeld:2000td}
  D.~Zeppenfeld, R.~Kinnunen, A.~Nikitenko and E.~Richter-Was,
  Phys.\ Rev.\  D {\bf 62} (2000) 013009
  [arXiv:hep-ph/0002036].
%
\bibitem{Klamke:2007cu}
  G.~Klamke and D.~Zeppenfeld,
  JHEP {\bf 0704} (2007) 052
  [arXiv:hep-ph/0703202].
%
\bibitem{Rainwater:1997dg}
  D.~L.~Rainwater and D.~Zeppenfeld,
  JHEP {\bf 9712} (1997) 005
  [arXiv:hep-ph/9712271].
%
\bibitem{Rainwater:1998kj}
  D.~L.~Rainwater, D.~Zeppenfeld and K.~Hagiwara,
  Phys.\ Rev.\  D {\bf 59} (1999) 014037
  [arXiv:hep-ph/9808468].
%
\bibitem{Plehn:1999xi}
  T.~Plehn, D.~L.~Rainwater and D.~Zeppenfeld,
  Phys.\ Rev.\  D {\bf 61} (2000) 093005
  [arXiv:hep-ph/9911385].
%
\bibitem{Eboli:2000ze}
  O.~J.~P.~Eboli and D.~Zeppenfeld,
  Phys.\ Lett.\  B {\bf 495} (2000) 147
  [arXiv:hep-ph/0009158].
%
\bibitem{Han:1992hr}
  T.~Han, G.~Valencia and S.~Willenbrock,
  Phys.\ Rev.\ Lett.\  {\bf 69} (1992) 3274
  [arXiv:hep-ph/9206246].
%
\bibitem{Figy:2003nv}
  T.~Figy, C.~Oleari and D.~Zeppenfeld,
  Phys.\ Rev.\  D {\bf 68} (2003) 073005
  [arXiv:hep-ph/0306109].
%
\bibitem{Asai:2004ws}
  S.~Asai {\it et al.},
  Eur.\ Phys.\ J.\  C {\bf 32S2} (2004) 19
  [arXiv:hep-ph/0402254].
%
\bibitem{Berger:2004pca}
  E.~L.~Berger and J.~M.~Campbell,
  Phys.\ Rev.\  D {\bf 70} (2004) 073011
  [arXiv:hep-ph/0403194].
%
\bibitem{Figy:2004pt}
  T.~Figy and D.~Zeppenfeld,
  Phys.\ Lett.\  B {\bf 591} (2004) 297
  [arXiv:hep-ph/0403297].
%
\bibitem{Ciccolini:2007jr}
  M.~Ciccolini, A.~Denner and S.~Dittmaier,
  Phys.\ Rev.\ Lett.\  {\bf 99} (2007) 161803
  [arXiv:0707.0381 [hep-ph]].
%
\bibitem{Ciccolini:2007ec}
  M.~Ciccolini, A.~Denner and S.~Dittmaier,
  Phys.\ Rev.\  D {\bf 77} (2008) 013002
  [arXiv:0710.4749 [hep-ph]].
%
\bibitem{DelDuca:2001fn}
  V.~Del Duca, W.~Kilgore, C.~Oleari, C.~Schmidt and D.~Zeppenfeld,
  Nucl.\ Phys.\  B {\bf 616} (2001) 367
  [arXiv:hep-ph/0108030].
%
\bibitem{Campbell:2006xx}
  J.~M.~Campbell, R.~K.~Ellis and G.~Zanderighi,
  JHEP {\bf 0610} (2006) 028
  [arXiv:hep-ph/0608194].
%
\bibitem{DelDuca:2006hk}
  V.~Del Duca {\it et al.},
  JHEP {\bf 0610} (2006) 016
  [arXiv:hep-ph/0608158].

%
\bibitem{Butterworth:2008iy}
  J.~M.~Butterworth, A.~R.~Davison, M.~Rubin and G.~P.~Salam,
  Phys.\ Rev.\ Lett.\  {\bf 100} (2008) 242001
  [arXiv:0802.2470 [hep-ph]].
%
\bibitem{Han:1991ia}
  T.~Han and S.~Willenbrock,
  Phys.\ Lett.\  B {\bf 273} (1991) 167.
%
\bibitem{Brein:2003wg}
  O.~Brein, A.~Djouadi and R.~Harlander,
  Phys.\ Lett.\  B {\bf 579}, 149 (2004)
  [arXiv:hep-ph/0307206].
%
\bibitem{Ciccolini:2003jy}
  M.~L.~Ciccolini, S.~Dittmaier and M.~Kramer,
  Phys.\ Rev.\  D {\bf 68} (2003) 073003
  [arXiv:hep-ph/0306234].
%
\bibitem{Brein:2004ue}
  O.~Brein, M.~Ciccolini, S.~Dittmaier, A.~Djouadi, R.~Harlander and M.~Kramer,
  arXiv:hep-ph/0402003.
%
%
%
\bibitem{Belyaev:2002ua}
  A.~Belyaev and L.~Reina,
  JHEP {\bf 0208} (2002) 041
  [arXiv:hep-ph/0205270].
%
\bibitem{Ball:2007zza}
  G.~L.~Bayatian {\it et al.}  [CMS Collaboration],
  J.\ Phys.\ G {\bf 34} (2007) 995.
%
%
\bibitem{Beenakker:2001rj}
  W.~Beenakker, S.~Dittmaier, M.~Kramer, B.~Plumper, M.~Spira and P.~M.~Zerwas,
  Phys.\ Rev.\ Lett.\  {\bf 87} (2001) 201805
  [arXiv:hep-ph/0107081].
%
\bibitem{Beenakker:2002nc}
  W.~Beenakker, S.~Dittmaier, M.~Kramer, B.~Plumper, M.~Spira and P.~M.~Zerwas,
  Nucl.\ Phys.\  B {\bf 653} (2003) 151
  [arXiv:hep-ph/0211352].
%
\bibitem{Dawson:2002tg}
  S.~Dawson, L.~H.~Orr, L.~Reina and D.~Wackeroth,
  Phys.\ Rev.\  D {\bf 67} (2003) 071503
  [arXiv:hep-ph/0211438].
%
\bibitem{Dawson:2003zu}
  S.~Dawson, C.~Jackson, L.~H.~Orr, L.~Reina and D.~Wackeroth,
  Phys.\ Rev.\  D {\bf 68} (2003) 034022
  [arXiv:hep-ph/0305087].
%
\bibitem{Bredenstein:2009aj}
  A.~Bredenstein, A.~Denner, S.~Dittmaier and S.~Pozzorini,
  arXiv:0905.0110 [hep-ph].
%
\bibitem{Bevilacqua:2009zn}
  G.~Bevilacqua, M.~Czakon, C.~G.~Papadopoulos, R.~Pittau and M.~Worek,
  arXiv:0907.4723 [hep-ph].
%
\end{thebibliography}
\end{document}